\input{aipcheck}

\documentclass[
    ,final            % use final for the camera ready runs
  ]
  {aipproc}
\usepackage{multirow}
\usepackage{amssymb}  
\usepackage{booktabs}
\usepackage{subfigure}  
  
\layoutstyle{8x11single}
\def\slashchar#1{\setbox0=\hbox{$#1$}
   \dimen0=\wd0 \setbox1=\hbox{/} \dimen1=\wd1
   \ifdim\dimen0>\dimen1 \rlap{\hbox to \dimen0{\hfil/\hfil}} #1
   \else  \rlap{\hbox to \dimen1{\hfil$#1$\hfil}} / \fi}
\def\p{\slashchar{p}}
\def\q{\slashchar{q}}
\def\D{\slashchar{D}}
\newcommand{\dr}[1]{\multirow{2}{*}{#1}}

\graphicspath{{./eps/}}

\begin{document}

\title{Weak Production of Strange Particles and $\eta$ Mesons off the Nucleon}


\author{M. Rafi Alam}{
  address={Department of Physics, Aligarh Muslim University, Aligarh-202 002, India}
}

\author{I. Ruiz Simo}{
  address={Departamento de F\'{\i}sica At\'omica, Molecular y Nuclear,
and Instituto de F\'{\i}sica Te\'orica y Computacional Carlos I,
Universidad de Granada, Granada 18071, Spain}
}

\author{L. Alvarez-Ruso}{
  address={Departamento de F\'\i sica Te\'orica and Instituto de F\'isica Corpuscular, Centro Mixto
Universidad de Valencia-CSIC, E-46071 Valencia, Spain}
}

\author{M. Sajjad Athar}{
  address={Department of Physics, Aligarh Muslim University, Aligarh-202 002, India}
}

\author{M. J. Vicente Vacas}{
  address={Departamento de F\'\i sica Te\'orica and Instituto de F\'isica Corpuscular, Centro Mixto
Universidad de Valencia-CSIC, E-46071 Valencia, Spain}
  % additional visiting address
}

\begin{abstract}
The strange particle production induced by (anti)neutrino off nucleon has been studied for $|\Delta S|=0$ and $|\Delta S|=1$ channels. 
 The reactions those we have considered are for the production of single kaon/antikaon, eta and associated particle production processes.
 We have developed a microscopical model based on the SU(3) chiral Lagrangian. The basic parameters of the model are
$f_\pi$, the pion decay constant, Cabibbo
angle, the proton and neutron magnetic moments and the axial
vector coupling constants for
the baryons octet. For antikaon production we have also included $\Sigma^*$(1385) resonance and for
eta production $S_{11}$(1535) and $S_{11}$(1650) resonances are included.
\end{abstract}

\maketitle

%%%%%%%%%%%%%%%%%%%%%%%%%%%%%%%%%%%%%%%%%%%%
%% MAINMATTER
%%%%%%%%%%%%%%%%%%%%%%%%%%%%%%%%%%%%%%%%%%%%
\section{Introduction}

Neutrino physics has become one of the important field of intense theoretical and 
experimental studies. 
In spite of huge efforts made in the last few decades to understand the 
nature of this elusive particle, there are still many unanswered questions like 
the absolute masses of neutrinos, CP violation in the lepton sector, etc. 
It has been established that neutrinos do oscillate and efforts are being made to 
determine the neutrino oscillation parameters better known as elements of PMNS matrix. 
Some of these oscillation parameters are sensitive to $\sim$1GeV of neutrino energies. 
For this a large number of experiments like 
T2K~\cite{Abe:2011ks}, MiniBooNE~\cite{AguilarArevalo:2007it}, MINER$\nu$A~\cite{Solomey:2005rs},
MINOS~\cite{Evans:2013pka}, No$\nu$A~\cite{Ayres:2004js}, LBNE~\cite{Adams:2013qkq}, INO~\cite{Athar:2006yb}
are going on or planned. 
Since the neutrino-nucleon cross sections are  very small, therefore, all these experiments 
are done using nuclear targets like $^{12}C$, $^{16}O$, $^{40}Ar$, $^{56}Fe$,  $^{208}Pb$ etc. 
In the few GeV energy region the contribution to the cross section comes from 
quasielastic, inelastic(like one pion, multipion, single kaon, single hyperon, associated 
particle production etc.) as well as from deep inelastic scattering processes. 
In the last two decades lots of efforts have been made to understand quasielastic 
scattering and one pion production processes. 
However, for the precise determination of neutrino oscillation parameters, it has been 
realized that the study of other inelastic processes like multi pion, single kaon, single hyperon, associated particle 
production, eta meson production etc. are also required to reduce the 
systematic uncertainties. 
The recent cross section measurements are available mostly for $\Delta S = 0$
processes in nonstrange sector with pions and/or nucleons in the final state. 
There are very few works available where
(anti)neutrino induced strange particle production
have been studied~\cite{Barish:1974ye,Barish:1978pj,Baker:1981tx,Baker:1981su}. 
However, with the presence of high intensity neutrino and antineutrino beam 
it is now possible to study reaction channels that involve the strange quark(s)
and give us an opportunity to test the SU(3) flavor symmetry.

The strange particles are produced by both $|\Delta S| =0 $ and $|\Delta S| = 1 $ processes. 
At the (anti)neutrino energies of $\sim$1GeV it is the single hyperon($Y$) or single kaon($K/\bar K$) that are produced by 
$|\Delta S| = 1 $ reaction mechanism while $\eta$ meson and associated production of kaon accompanied by a hyperon($K+Y$)
are produced by the $|\Delta S| = 0 $ mechanism. 
The reaction cross sections are smaller than the pion production due to Cabibbo suppression in $|\Delta S| = 1 $ process and due to the low  phase space 
for $|\Delta S| = 0$ process. These processes are also important to estimate the background for the experiments 
performing for nucleon decay searches.
Furthermore, these processes would also help in determining the various transition form factors.

In this work we have studied  neutrino/antineutrino
induced $1K,YK$ and $\eta$ production processes, where $K$ stands for kaon and $Y$ for hyperon, 
and the various processes are given as:
 
 \begin{enumerate}
  \item Single Kaon Production
  
\begin{minipage}{0.45\textwidth}
\begin{eqnarray*}
\nu_l + p &\rightarrow & l^- + K^+ + p  \nonumber\\
\nu_l + n &\rightarrow & l^- + K^0 + p  \nonumber\\
\nu_l + n &\rightarrow & l^- + K^+ + n   
\end{eqnarray*}
\end{minipage}
\begin{minipage}{0.45\textwidth}
\begin{eqnarray}\label{Eq1:single_Kprod}
\bar \nu_l + p &\rightarrow & l^+ + K^- + p  \nonumber\\
\bar \nu_l + p &\rightarrow & l^+ +\bar K^0 + n  \nonumber\\
\bar \nu_l + n &\rightarrow & l^+ + K^- + n ,  
\end{eqnarray}
\end{minipage}\\

\item Associated Kaon Production

\begin{minipage}{0.45\textwidth}
\begin{eqnarray*}
\nu_l + n ~  &\rightarrow&~  l^- + \Lambda^0 +   K^+  \nonumber \\
\nu_l + n ~  &\rightarrow&~  l^- + \Sigma^0 +  K^+ \nonumber \\
\nu_l + n ~  &\rightarrow&~  l^- + \Sigma^+ + K^0 \nonumber \\
\nu_l + p ~  &\rightarrow&~  l^- + \Sigma^+ + K^+ 
\end{eqnarray*}
\end{minipage}
\begin{minipage}{0.45\textwidth}
\begin{eqnarray}\label{Eq1:ass_prod}
\bar\nu_l + p ~  &\rightarrow&~  l^+ + \Lambda^0 + K^0 \nonumber \\
\bar\nu_l + p ~  &\rightarrow&~  l^+ + \Sigma^0 + K^0 \nonumber \\ 
\bar\nu_l + p ~  &\rightarrow&~  l^+ + \Sigma^- + K^+ \nonumber \\ 
\bar\nu_l + n ~  &\rightarrow&~  l^+ + \Sigma^- + K^0 ,  
\end{eqnarray}
\end{minipage}\\

\item Weak eta meson production

\begin{equation}
 \nu_l + n \rightarrow l^-  + \eta + p   \quad \quad \rm{and}  \quad \quad   \quad \quad 
\bar \nu_l + p  \rightarrow l^+  + \eta + n.  \label{eta}
\end{equation}

\end{enumerate}

The plan of the presentation is as follows. First we present the formalism in brief then we will discuss the results and 
finally we conclude our findings.
 
% 
% %%%%%%%%%%%%%%%%%%%%%%%%%%%%%%%%%%%%%%%%%%%%%%%%
%% Formalism
%%%%%%%%%%%%%%%%%%%%%%%%%%%%%%%%%%%%%%%%%%%%%%%%
%  \section{interaction channels}
\section{Formalism}\label{sec:sec1}
The general expression for the scattering cross-section in Lab frame may be written as,
\begin{equation}\label{eq:kinc}
d^{9}\sigma = \frac{(2\pi)^{4}}{4 M E} \prod_{f=1}^{n} \frac{d{\vec k_f} }{ 2 k^0_f (2\pi)^{3})} 
 \delta^{4}( k_i - k_f )\bar\Sigma\Sigma | \mathcal M |^2,
\end{equation}
where  $ \vec{k_f}$  is the 3-momenta of the incoming and/or outgoing leptons in the lab frame with energy $k^0_f$.
$E$ is the energy of incoming neutrino beam, $M$ is the nucleon mass,
$ \bar\Sigma\Sigma | \mathcal M |^2  $ is the square of the transition amplitude matrix 
element averaged(summed) over the spins of the initial(final) state. 
At low energies, this amplitude can be written as
\begin{equation}
 \mathcal M = \frac{G_F}{\sqrt{2}}\, j_\mu^{(L)} J^{\mu\,{(H)}}=\frac{g}{2\sqrt{2}}j_\mu^{(L)} \frac{1}{M_W^2}
\frac{g}{2\sqrt{2}}J^{\mu\,{(H)}},
\end{equation}
 where $j_\mu^{(L)}$ and $  J^{\mu\,(H)}$ are the leptonic and hadronic currents respectively, 
$G_F=\sqrt{2} \frac{g^2}{8 M^2_W}=1.16639(1)\times 10^{-5}\,\mbox{GeV}^{-2}$ is the Fermi constant and 
$g$ is the weak gauge coupling.
The Standard Model Lagrangian for leptonic current is given by,
\begin{equation}
{\cal L}=-\frac{g}{2\sqrt{2}}\left[{ W}^+_\mu\bar{\nu}_l
\gamma^\mu(1-\gamma_5)l+{ W}^-_\mu\bar{l}\gamma^\mu
(1-\gamma_5)\nu_l\right],
\end{equation}
where,  $W$ boson couples to the leptons. 
In the neutrino energy of $\sim$1GeV hadronic current may be obtained 
using the chiral perturbation theory~\cite{RafiAlam:2010kf,Alam:2011xq}.

To obtain the hadronic current, one starts with the  lowest-order SU(3) chiral
Lagrangian describing the pseudoscalar mesons in the presence of an external current~\cite{Scherer:2002tk}: 
\begin{equation}
\label{eq:lagM}
{\cal L}_M^{(2)}=\frac{f_\pi^2}{4}\mbox{Tr}[D_\mu U (D^\mu U)^\dagger]
+\frac{f_\pi^2}{4}\mbox{Tr}(\chi U^\dagger + U\chi^\dagger),
\end{equation}
where the parameter $f_\pi=92.4$MeV is the pion  decay constant, $U$ is the SU(3) representation of the meson fields given by
\begin{eqnarray}
U(x)&=&\exp\left(i\frac{\phi(x)}{f_\pi}\right),\nonumber\\
\phi(x)&=&
\left(\begin{array}{ccc}
\pi^0+\frac{1}{\sqrt{3}}\eta &\sqrt{2}\pi^+&\sqrt{2}K^+\\
\sqrt{2}\pi^-&-\pi^0+\frac{1}{\sqrt{3}}\eta&\sqrt{2}K^0\\
\sqrt{2}K^- &\sqrt{2}\bar{K}^0&-\frac{2}{\sqrt{3}}\eta
\end{array}\right),
\end{eqnarray}
and $D_\mu U$ is its covariant derivative
\begin{eqnarray}
D_\mu U&\equiv&\partial_\mu U -i r_\mu U+iU l_\mu\,.
\end{eqnarray}
Here, $l_\mu$ and $r_\mu$ corresponds to left and right handed currents, that for the charged current(CC) case are given by
\begin{equation}
r_\mu=0,\quad l_\mu=-\frac{g}{\sqrt{2}}
({W}^+_\mu T_+ + {W}^-_\mu T_-),
\end{equation}
with $W^\pm$ the $W$ boson fields and
$$
T_+=\left(\begin{array}{rrr}0&V_{ud}&V_{us}\\0&0&0\\0&0&0\end{array}\right);\quad
T_-=\left(\begin{array}{rrr}0&0&0\\V_{ud}&0&0\\V_{us}&0&0\end{array}\right).
$$
Here,  $V_{ij}$ are the elements of the 
Cabibbo-Kobayashi-Maskawa  matrix. 

Similarly, lowest-order  chiral Lagrangian for the baryon octet$(B)$ in the presence 
of an external current is written in terms of the SU(3) matrix~\cite{Scherer:2002tk}
\begin{equation}
B=
\left(\begin{array}{ccc}
\frac{1}{\sqrt{2}}\Sigma^0+\frac{1}{\sqrt{6}}\Lambda&\Sigma^+&p\\
\Sigma^-&-\frac{1}{\sqrt{2}}\Sigma^0+\frac{1}{\sqrt{6}}\Lambda&n\\
\Xi^-&\Xi^0&-\frac{2}{\sqrt{6}}\Lambda
\end{array}\right)
\end{equation}
as 
\begin{equation}
{\cal L}^{(1)}_{MB}=\mbox{Tr}\left[\bar{B}\left(i\D
-M\right)B\right]
-\frac{D}{2}\mbox{Tr}\left(\bar{B}\gamma^\mu\gamma_5\{u_\mu,B\}\right)
-\frac{F}{2}\mbox{Tr}\left(\bar{B}\gamma^\mu\gamma_5[u_\mu,B]\right),
\end{equation}\label{eq:lagB}

where $M$ denotes the mass of the baryon octet, and the parameters $D=0.804$ and $F=0.463$
are determined from the baryon semileptonic decays~\cite{Cabibbo:2003cu}.
The covariant derivative of baryon octet $(B)$ is given by
\begin{equation}
D_\mu B=\partial_\mu B +[\Gamma_\mu,B],
\end{equation}
with
\begin{equation}
\Gamma_\mu=\frac{1}{2}\left[u^\dagger(\partial_\mu-ir_\mu)u
+u(\partial_\mu-il_\mu)u^\dagger\right],
\end{equation}
where  we have introduced $u^2=U$ with,
\begin{equation}
u_\mu= i\left[u^\dagger(\partial_\mu-i r_\mu)u-u(\partial_\mu-i
l_\mu)u^\dagger\right].
\end{equation}
The next order meson baryon Lagrangian contains many new terms, which has been not considered in the present calculations and their 
contribution would be small. We have used the prescription given by Cabibbo.~\cite{RafiAlam:2010kf,Alam:2011xq,Cabibbo:2003cu}. 
 To include the higher order terms in the Lagrangian in this case, the coupling constants are fully determined by the proton and neutron anomalous magnetic moments.
Now we shall present the amplitudes corresponding to the 
hadronic current for different processes given in Eqs.~\ref{Eq1:single_Kprod}-\ref{eta}.

\subsection{Single Kaon Production}
In the neutrino-nucleon scattering process the first process that produces strange particle
in the  inelastic region  is the $\Delta S =1 $ kaon/antikaon $(K/\bar K)$ production. 
These processes are Cabibbo suppressed, however, in the absence of any competing process that 
could produce kaon and/or antikaon in the neutrino energies below 1.5GeV, they are the 
dominant source of $K/\bar K$ production. 
For example, the threshold for the $ \nu(\bar \nu) N \rightarrow l^-(l^+) N K(\bar K) $ 
is about 750MeV, as compared to 
the next $K$-production mechanism $ \nu/\bar \nu N \rightarrow l^\pm Y  K $ 
and $\bar K$-production mechanism $ \nu/\bar \nu N \rightarrow l^\pm N   K  \bar K $ 
for which the threshold is  1.2GeV and 1.8GeV respectively. 
Therefore, the study of these processes could be important in giving ansatz to some important 
strangeness physics at low energies.

\begin{figure}\label{fig:feyn_sing_K}
 \includegraphics[width=10cm,height=4cm]{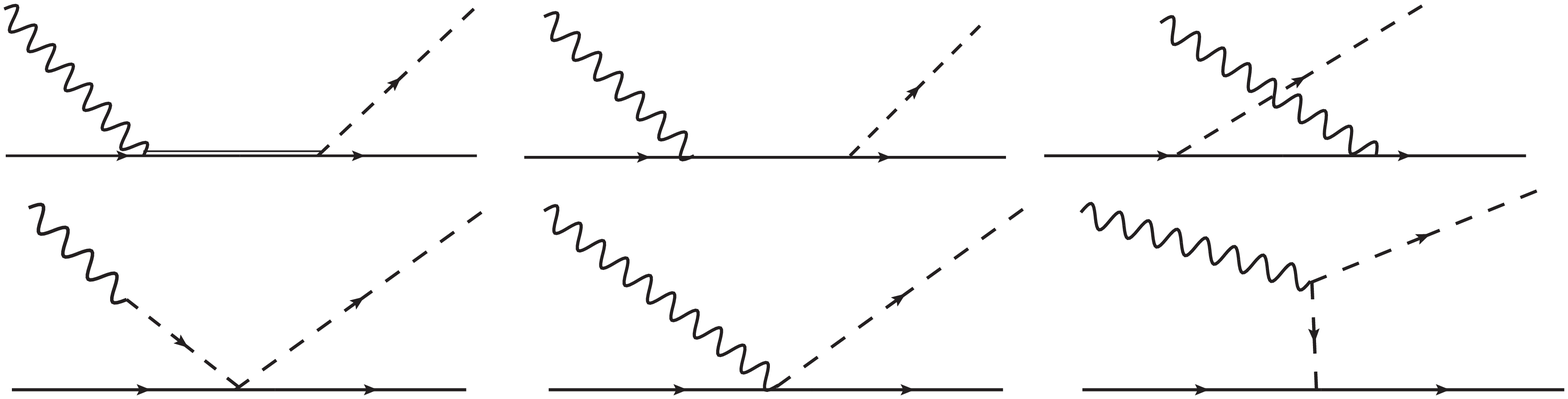}
 \caption{Feynman diagrams for $\Delta S =1 $ kaon production. Top row right to left 
  $Y^\ast$ resonance, direct and cross hyperon pole. 
  Second row consists of diagrams (right to left) kaon pole, contact and kaon in flight.}
\end{figure}

The amplitude for the hadronic current gets contribution from various Feynman diagrams shown in 
Fig.~\ref{fig:feyn_sing_K} viz. direct$(Y)$, cross hyperon pole$(CrY)$, contact$(CT)$ diagram, pion pole$(KP)$,and 
pion/eta in flight$(\pi, \eta)$. We have also included $\Sigma^*$(1385) resonance for antikaon production. 
For single kaon production the amplitudes are obtained as,
\begin{eqnarray}\label{eq:1k_amp}
J^\mu \arrowvert_{CT} &=&i A_{CT} V_{us} \frac{ \sqrt{2}}{2 f_\pi}  \bar N(p^\prime) 
\; (\gamma^\mu + B_{CT} \; \gamma^\mu \gamma_5 ) \; N(p) \nonumber \\
j^{\mu}\big|_{Cr\Sigma} &=& i A_{Cr\Sigma} V_{us} (D-F) \frac{\sqrt{2}}{2 f_\pi} \bar N(p^\prime)
 \left( \gamma^\mu 
+{i\frac{\mu_p+2\mu_n}{2M}\sigma^{\mu\nu}q_\nu}
+ (D-F)(\gamma^\mu-{\frac{q^\mu}{q^2-M_k^2}\q } )\gamma^5 \right)
\frac{\p - \p_k + M_\Sigma}{( p -  p_k)^2 -M_\Sigma^2} \p_k \gamma^5  N(p)  , \nonumber \\
j^{\mu}\big|_{Cr\Lambda}&=&i A_{Cr\Lambda} V_{us} (D+3F) \frac{\sqrt{2}}{4 f_\pi} \bar N(p^\prime)\left( \gamma^\mu
 +{i\frac{\mu_p}{2M}\sigma^{\mu\nu}q_\nu}
-\frac{D+3F}{3} (\gamma^\mu -{\frac{q^\mu}{q^2-M_k^2}\q } )\gamma^5  \right) 
 \frac{\p - \p_k +M_\Lambda}{( p -  p_k)^2 -M_\Lambda^2}   \p_k \gamma^5 N(p) , \nonumber \\
J^\mu \arrowvert_{\Sigma} &=&i A_{\Sigma} (D-F) V_{us} \frac{ \sqrt{2}}{2 f_\pi} 
\bar N(p^\prime) p_k\hspace{-.9em}/ \; \gamma_5  \frac{ p\hspace{-.5em}/ +
  q\hspace{-.5em}/ + M_\Sigma}
  {( p +  q)^2 -M_\Sigma^2} \left(\gamma^\mu + 
  { i  \frac{(\mu_p + 2\mu_n)}{2 M} \sigma^{\mu \nu} q_\nu} +
  (D-F) \left\{ \gamma^\mu 
  - {\frac{q^\mu}{ q^2-{M_k}^2 } q\hspace{-.5em}/} \right\} \gamma^5 \right) N(p) \nonumber \\
J^\mu \arrowvert_{\Lambda} &=&  i A_{\Lambda} V_{us} (D+3F)  \frac{1} {2 \sqrt{2} f_\pi} 
\bar N(p^\prime) p_k\hspace{-.9em}/ \; \gamma^5 \frac{ p\hspace{-.5em}/ +
  q\hspace{-.5em}/ +M_\Lambda}
  {( p +  q)^2 -M_\Lambda^2} \left(\gamma^\mu +
    {  i \frac{\mu_p}{2 M}  \sigma^{\mu \nu} q_\nu } -
  \frac{(D + 3 F)}{3} \left\lbrace \gamma^\mu   -  {  \frac{q^\mu }{ q^2-{M_k}^2 } q\hspace{-.5em}/}
 \right\rbrace \gamma^5 \right) N(p) \nonumber \\
J^\mu \arrowvert_{KP}&=& i A_{KP} V_{us} \frac{\sqrt{2}}{2 f_\pi}  \bar N(p^\prime) 
q\hspace{-.5em}/ \; N(p) \frac{q^\mu}{q^2-M_k^2}  \nonumber  \\
J^\mu \arrowvert_{\pi} &=& iA_{\pi } \frac{M\sqrt{2}}{2 f_\pi}  V_{us} 
(D + F)\frac{ 2 {p_k}^\mu -q^\mu}{(q-p_k)^2 - {m_\pi}^2} \bar N(p^\prime)  \gamma_5  N(p) \nonumber \\
J^\mu \arrowvert_{\eta} &=&i A_{\eta } \frac{M\sqrt{2}}{2 f_\pi}  V_{us} 
(D - 3 F)\frac{2 {p_k}^\mu - q^\mu}{(q-p_k)^2 - {m_\eta}^2} \bar N(p^\prime) 
        \gamma_5 N(p) \nonumber \\
J^\mu \arrowvert_{\Sigma^*} &=&- i A_{\Sigma^*} \frac{\cal C}{ f_\pi } \frac{1}{\sqrt{6}} \; V_{us} \;  
  \frac{p_k^\lambda}{P^2 - M_{\Sigma^*}^2 + i \Gamma_{\Sigma^*} M_{\Sigma^*}}\; 
\bar N(p^\prime) P_{RS_{\lambda \rho}} ( \Gamma_V^{\rho \mu} +\Gamma_A^{\rho \mu} ) N(p),
\end{eqnarray}
where, $q=k-k^\prime$ is the four momentum transfer, $P=p+q$ is the momentum carried by the resonance and 
$P^{\mu \nu}_{RS}$ is the Rarita-Schwinger projection operator given by
\begin{equation}
P^{\mu\nu}_{RS}(P)= \sum_{spins} \psi^{\mu} \bar \psi^{\nu} =- (\slashchar{P} + M_{\Sigma^*}) \left [ g^{\mu\nu}-
  \frac13 \gamma^\mu\gamma^\nu-\frac23\frac{P^\mu
  P^\nu}{M_{\Sigma^*}^2}+ \frac13\frac{P^\mu
  \gamma^\nu-P^\nu \gamma^\mu }{M_{\Sigma^*}}\right],
\label{eq:rarita_prop}
\end{equation}
with $M_{\Sigma^*}$ the resonance mass  and $\psi^{\mu}$ 
 the Rarita-Schwinger spinor. 
 The $\Sigma^\ast B \phi$($\phi \equiv \rm{meson}, \; B \equiv \rm{baryon}$) coupling$(\cal C)$
 is obtained from the on-shell $\Sigma^*$ width
\begin{eqnarray}
 \Gamma_{\Sigma^*}&=&\Gamma_{\Sigma^*\rightarrow \Lambda \pi} 
+ \Gamma_{\Sigma^*\rightarrow \Sigma \pi}+ \Gamma_{\Sigma^*\rightarrow N \bar{K}}\; ,
\label{eq:width}
\end{eqnarray}
where
\begin{eqnarray}
 \Gamma_{\Sigma^* \rightarrow Y,\, \phi }&=&\frac{C_Y}{192\pi}\left(\frac{\cal C}{f_\pi}\right)^2
\frac{(W+M_Y)^2-m^2}{W^5}\lambda^{3/2}(W^2,M_Y^2,m^2)  \Theta(W-M_Y-m).
\end{eqnarray}
Here, $m,\, M_Y$ are the masses of the emitted meson and baryon,
 $\lambda(x,y,z)=(x-y-z)^2-4yz$ and $\Theta$ is the 
step function. The factor $C_Y$   is 1 for $\Lambda$ and $\frac23$ for $N $ and $\Sigma$. 

The  $W^- N \rightarrow \Sigma^*$ vertex may be written in terms of 
a vector and an axial-vector piece given by~\cite{Alam:2011xq}
\begin{eqnarray} \label{eq:delta_amp}
\langle \Sigma^{*}; P= p+q\, | V^\mu | N;
p \rangle &=& V_{us} \bar\psi_\alpha(\vec{P} ) \Gamma^{\alpha\mu}_V \left(p,q \right)
u(\vec{p}\,), \nonumber \\
\langle \Sigma^{*}; P= p+q\, | A^\mu | N;
p \rangle &=& V_{us} \bar \psi_\alpha(\vec{p} ) \Gamma^{\alpha\mu}_A \left(p,q \right)
u(\vec{p}\,),
\end{eqnarray}
where
\begin{eqnarray}
\Gamma^{\alpha\mu}_V (p,q) &=&
\left [ \frac{C_3^V}{M}\left(g^{\alpha\mu} \slashchar{q}-
q^\alpha\gamma^\mu\right) + \frac{C_4^V}{M^2} \left(g^{\alpha\mu}
q\cdot P- q^\alpha P^\mu\right)  + \frac{C_5^V}{M^2} \left(g^{\alpha\mu}
q\cdot p- q^\alpha p^\mu\right) + C_6^V g^{\mu\alpha}
\right ]\gamma_5 \nonumber\\
\Gamma^{\alpha\mu}_A (p,q) &=& \left [ \frac{C_3^A}{M}\left(g^{\alpha\mu} \slashchar{q}-
q^\alpha\gamma^\mu\right) + \frac{C^A_4}{M^2} \left(g^{\alpha\mu}
q\cdot P- q^\alpha P^\mu\right)  + C_5^A g^{\alpha\mu} + \frac{C_6^A}{M^2} q^\mu q^\alpha
\right ]. \label{eq:del_ffs}
 \end{eqnarray}

\begin{table}
%  \begin{center}|l|c|c|c|c|c|c|c|c|c|c|
\begin{tabular}{lcccccccccc} \hline \hline
	    Process   	  	       &$B_{CT}$&$A_{CT}$& $A_{\Sigma}$ &$A_{\Lambda}$&$A_{Cr\Sigma}$&$A_{Cr\Lambda}$&$A_{KP}$& $A_{\pi }$ & $A_{\eta }$ & $ A_{\Sigma^*} $ \\\hline		
$ \nu n \rightarrow l^-  K^+ n    $&    D-F &   -1   &    0 	&    0        &   -1	     &    0	    &-1       &    -1	   &   -1	 &  0		    \\  	     
$ \nu p \rightarrow l^-  K^+ p    $&     -F &   -2	 &    0 	&    0        &$-\frac{1}{2}$&    1	    &-2       &    1	   &   -1	 &  0		    \\  	     
$ \nu n \rightarrow l^-  K^0 p    $&   -D-F &   -1	 &    0 	&    0        & $\frac{1}{2}$&    1	    &-1       &    2	   &   0	 &  0		    \\ 	       
$ \bar \nu n \rightarrow l^+  K^- n        $&    D-F &    1   &    -1	&   0	      &    0	     &    0	    &-1       &    1	   &   1	 &  2		    \\		
$ \bar \nu p \rightarrow l^+  K^- p        $&     -F &   2	 &$-\frac{1}{2}$&   1	      &    0	     &    0	    &-2       &   -1	   &   1	 &  1		    \\		
$ \bar \nu p \rightarrow l^+ \bar K^0 n    $&   -D-F &   1	 & $\frac{1}{2}$&   1	      &    0	     &    0	    &-1       &   -2	   &   0	 &  -1  	    \\		
\hline \hline   
 \end{tabular}
\caption{Constant factors  appearing in the hadronic current}\label{tb:1k_amp}
%  \end{center}
\end{table}
The form factors $C_i^V,C_i^A,\;(i=3,4,5,6)$ are obtained using SU(3) symmetry and 
discussed in detail in Ref.~\cite{Alam:2011xq}.

\subsection{Associated production of strange particles}

At intermediate neutrino energies(1-2 GeV) kaon 
may also get produced in  strangeness conserving $|\Delta S| = 0$ processes 
where a $\Lambda$ or $\Sigma$ are produced with kaon. 
However, as the hyperon mass is slightly larger than the 
nucleon mass, kinematically near threshold the production of 
kaons are favored by the $|\Delta S| =1 $ processes. 

\begin{figure}\label{fig:feynman_ass}
 \includegraphics[width=14cm,height=6cm]{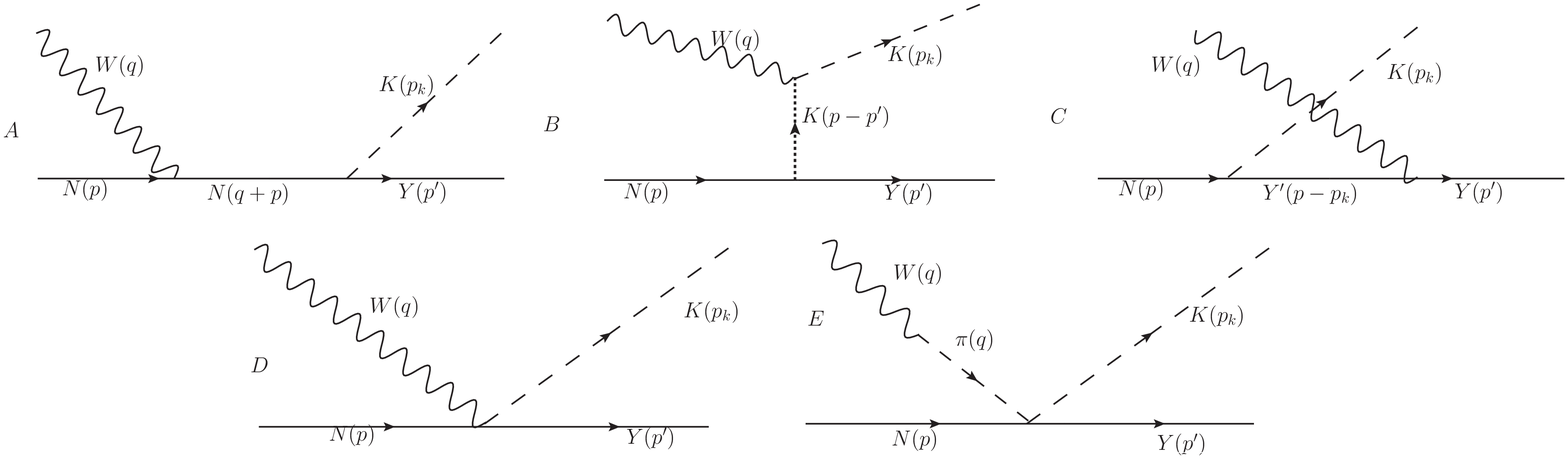}
 \caption{Feynman diagrams for $\Delta S$=0 kaon production. $(A)$direct nucleon pole(s),$~(B)$kaon pole(t),
 $~(C)$cross hyperon pole(u),$~(D)$Contact diagrams(CT),$~(E)$Pion in flight($\pi$F) }
 \end{figure}

The Feynman diagrams that contribute to the $\Delta S = 0 $ kaon production processes
are depicted in Fig.~\ref{fig:feynman_ass}.  
To get the coupling of the amplitudes for the above diagrams we used the same prescription as 
used in the case of $\Delta S = 1 $ kaon production processes.  
However, in this case we generalize the nucleon and/or hyperon pole diagrams to incorporate form factors
at the weak vertices, whereas a dipole form has been taken for the $|\Delta S| = 1$ single kaon production processes. 
The matrix element corresponding to the
hyperon nucleon transition may be written as,
\begin{equation}
 J^\mu = \langle Y (k^\prime) | V^\mu - A^\mu | N(k) \rangle 
\end{equation}
The amplitude thus obtained corresponding to the diagrams shown in Fig.~\ref{fig:feynman_ass} and
are written as

\begin{table}
%\begin{center}
\begin{tabular}{lccccccc} \toprule 
Process 	                                & $A_{CT}$	            &$B_{CT}$		      &     $A_{SY}$  	             & $A_{U\Sigma}$                 & $A_{U\Lambda}$      & $A_{TY}$                     &  $A_{\pi }$ \\ \toprule 
$\bar\nu_l  p \rightarrow l^+  \Lambda  K^0 $ &\dr{$-\sqrt{\frac32}$}       &\dr{$\frac{-1}{3}(D+3F)$}&\dr{$\frac{-1}{\sqrt6}(D+3F)$}& \dr{ $-\sqrt{\frac23}(D-F)$ } &  \dr{  0}           &\dr{$\frac{-1}{\sqrt6}(D+3F)$}& \dr{$\sqrt{\frac32}$} \\
$\nu_l n \rightarrow l^- \Lambda  K^+  $ & & & & & & & \\ \midrule
$ \bar\nu_l  p \rightarrow l^+  \Sigma^0  K^0 $ &\dr{$\mp \frac{1}{\sqrt2}$}&\dr{$D-F$}	              &\dr{$\mp\frac{1}{\sqrt2}(D-F)$}& \dr{$\mp \sqrt2 (D-F)$ }     &   \dr{0}            &\dr{$\pm\frac{1}{\sqrt2}(D-F)$}& \dr{$\pm\frac{1}{\sqrt2}$ }  \\ 
$\nu_l n \rightarrow l^-\Sigma^0 K^+ $  & & & & & & & \\  \midrule %\midrule\specialrule{2.5pt}{1pt}{1pt}
$ \bar\nu_l  p \rightarrow l^+  \Sigma^- K^+ $  & \dr{0}		    &\dr{0}  		      &   \dr{ $D-F$	}	     & \dr{$D-F$}                    & \dr{$\frac13(D+3F)$}& \dr{0}         	          & \dr{ 0}	   \\
$\nu_l n \rightarrow l^-\Sigma^+ K^0 $  & & & & & & & \\  \midrule
$ \bar\nu_l  n \rightarrow l^+  \Sigma^-  K^0 $ & \dr{$-1$}		    & \dr{$D-F$}	      &\dr{0}  		             & \dr{$F-D$}	             &\dr{$\frac13(D+3F)$} &   \dr{$D-F$ }     	          & \dr{ $1$ } \\
$\nu_l p \rightarrow l^-\Sigma^+ K^+ $  & & & & & & & \\ 
\bottomrule
\end{tabular}
\caption{Constant factors  appearing in the hadronic current. The upper sign corresponds to the 
processes with $\bar \nu$ }\label{tb:currents}
%\end{center}
\end{table}

\begin{eqnarray}\label{eq:ass_amp}
j^\mu \arrowvert_{s} &=& i A_{SY} V_{ud} \frac{\sqrt2}{2f_\pi} \; \bar u_Y (p^\prime) \slashchar p_k \gamma^5 
	    \frac{\slashchar p + \slashchar q + M}{(p+q)^2-M^2} 
	   {  {\cal H}^\mu } u_N(p) \nonumber \\
j^\mu \arrowvert_{u} &=& i A_{UY} V_{ud} \frac{\sqrt2}{2f_\pi} \; \bar u_Y (p^\prime) { {\cal H}^\mu }
	      \frac{\slashchar p - \slashchar p_k + M_{Y^\prime}}{(p - p_k)^2-M_{Y^\prime}^2}  \slashchar p_k \gamma^5 u_N (p) \nonumber \\
j^\mu \arrowvert_{t} &=& i A_{TY} V_{ud} \frac{\sqrt2}{2f_\pi} (M+M_Y) \; \bar u_Y (p^\prime) \gamma_5 \; u_N (p) \;\;
	    \frac{q^\mu - 2 p_k^\mu}{(p-p^\prime)^2-m_k^2}\nonumber \\
j^\mu \arrowvert_{CT} &=& i A_{CT} V_{ud} \frac{\sqrt2}{2f_\pi} \; \bar u_Y (p^\prime) 
		    \left( \gamma^\mu + B_{CT} \; \gamma^\mu  \gamma^5 \right) u_N (p) \nonumber \\
j^\mu \arrowvert_{\pi F} &=& i A_{\pi} V_{ud} \frac{\sqrt2}{4f_\pi} \; \bar u_Y (p^\prime) 
	      (\slashchar q + \slashchar p_k) u_N(p) \frac{q^\mu}{q^2-m_\pi^2} \nonumber \\
{\cal H}^\mu &=& f_1^V \gamma^\mu + i \frac{f_2^V}{2M} \sigma^{\mu \nu} q_\nu
	    - f_A \left(\gamma^\mu - { \frac{\slashchar q q^\mu}{q^2 -m_\pi^2} } \right)\gamma^5,
\end{eqnarray}
where ${\cal H}^\mu$ is the transition current for $ Y \leftrightarrows Y^\prime $ with 
$Y= Y^\prime \equiv $ Nucleon and/or Hyperon. 
The constant factors$(A_i)$ appearing in Eq.~\ref{eq:ass_amp} are summarized in Table~\ref{tb:currents}.

The form factors $f_1^V,f_2^V$ are related to the proton$(f_{1,2}^{p})$ and neutron$(f_{1,2}^{n})$ 
transition form factors, the details of which are given in Ref.~\cite{Alam:2014bya} and are tabulated in Tab~\ref{tab:ass_ff}.
The $f_{1,2}^{p,n}$ are parameterized with the help of electromagnetic Sach's form factors~\cite{Singh:2006}.  
The axial vector coupling $f_A(Q^2) = \frac{f_A(0)}{(1+Q^2/M_A^2)^2}$ is parameterized with  dipole form with 
the axial mass $M_A=1.05$GeV and  $f_A(0)  = 1.26$.

\begin{figure}
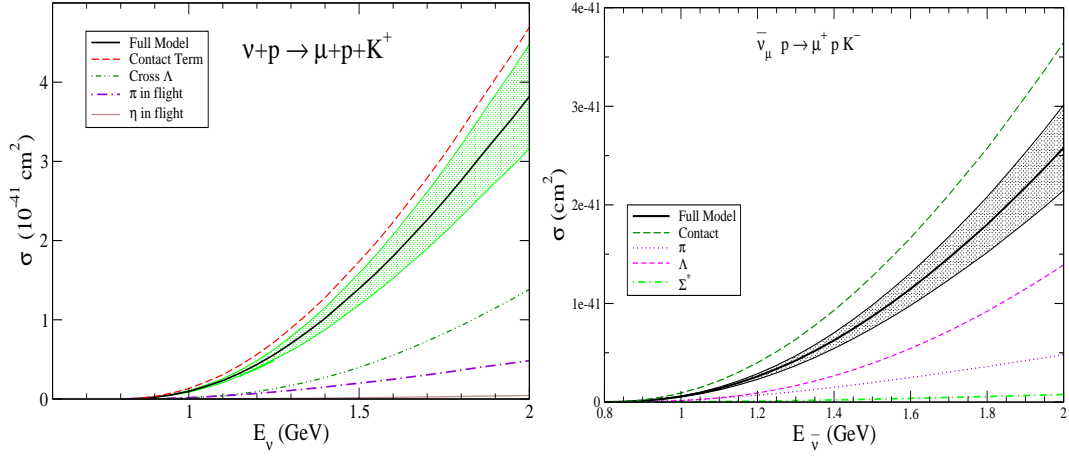
\label{fig:pp_singleK}
 \includegraphics[width=7cm,height=6cm]{pp2.eps}
 \includegraphics[width=7cm,height=6cm]{PP.eps}
 \caption{Cross section for $\nu/\bar{\nu}$ induced $K^+~\&~ K^-$production processes.}
\end{figure}

 \begin{table}
%\begin{ruledtabular}
% use packages: array
\renewcommand{\arraystretch}{2}
\begin{tabular}{llccc}
\hline\hline
Sl. No. &Weak transition & $f_1^V(Q^2)$& $f_2^V(Q^2)$& $f_A(Q^2)$ \\ \hline \hline
1&${p} \rightarrow {n}$ & $f^{p}_1(Q^2) - f^{n}_1(Q^2)$ & $f^{p}_2(Q^2) - f^{n}_2(Q^2)$ & $g_A(Q^2)$ \\ \hline
\dr{2}&${p} \rightarrow \Lambda $ & $-\sqrt{\frac{3}{2}} f^{p}_1(Q^2)$ & $-\sqrt{\frac{3}{2}} f^{p}_2(Q^2)$ & $-\sqrt{\frac{1}{6}}\frac{3F + D}{F + D}g_A(Q^2)$\\ 
      &${n} \rightarrow \Lambda $ & $-\sqrt{\frac{3}{2}} f^{p}_1(Q^2)$ & $-\sqrt{\frac{3}{2}} f^{p}_2(Q^2)$ & $-\sqrt{\frac{1}{6}}\frac{3F + D}{F + D}g_A(Q^2)$\\ \hline
3&$\Sigma^{\pm}  \rightarrow \Lambda$ & $-\sqrt{\frac{3}{2}}f^{n}_1(Q^2)$ & $-\sqrt{\frac{3}{2}}f^{n}_2(Q^2)$ & $\sqrt{\frac{2}{3}}\frac{D}{F + D} g_A(Q^2)$ \\ \hline
4&$\Sigma^{\pm} \rightarrow \Sigma^{0}$ & $\mp\frac{1}{\sqrt{2}}[2f^{p}_1(Q^2) + f^{n}_1(Q^2)] $ & $\mp \frac{1}{\sqrt{2}}[2f^{p}_2(Q^2) + f^{n}_2(Q^2)]$ & $\mp\sqrt{2}\frac{F}{F + D}g_A(Q^2)$\\ \hline
\dr{5}&$p \rightarrow \Sigma^{0}$ & $-\frac{1}{\sqrt{2}}[f^{p}_1(Q^2) + 2f^{n}_1(Q^2)] $ & $-\frac{1}{\sqrt{2}}[f^{p}_2(Q^2) + 2f^{n}_2(Q^2)] $ & $\frac{1}{\sqrt{2}}\frac{D-F}{F + D}g_A(Q^2)$\\
&$n\rightarrow \Sigma^{0}$ & $\frac{1}{\sqrt{2}}[f^{p}_1(Q^2) + 2f^{n}_1(Q^2)]$&$\frac{1}{\sqrt{2}}[f^{p}_2(Q^2) + 2f^{n}_2(Q^2)]$   &  $-\frac{1}{\sqrt{2}}\frac{D-F}{F + D}g_A(Q^2)$\\ \hline
\dr{6}&$n\rightarrow \Sigma^{-}$ &  $-f^{p}_1(Q^2) - 2f^{n}_1(Q^2)$ & $-f^{p}_2(Q^2) - 2f^{n}_2(Q^2)$  & $\frac{D-F}{F + D}g_A(Q^2)$\\ 
&$p\rightarrow \Sigma^{+}$ &  $-f^{p}_1(Q^2) - 2f^{n}_1(Q^2)$ & $-f^{p}_2(Q^2) - 2f^{n}_2(Q^2)$  & $\frac{D-F}{F + D}g_A(Q^2)$\\  \hline \hline
\end{tabular}
\label{tab:ass_ff}
%\end{ruledtabular}
\caption{Isovector ($f^V_1 , f^V_2$ and axial$(f_A)$ transition form factors.)}
\end{table}

\begin{figure}
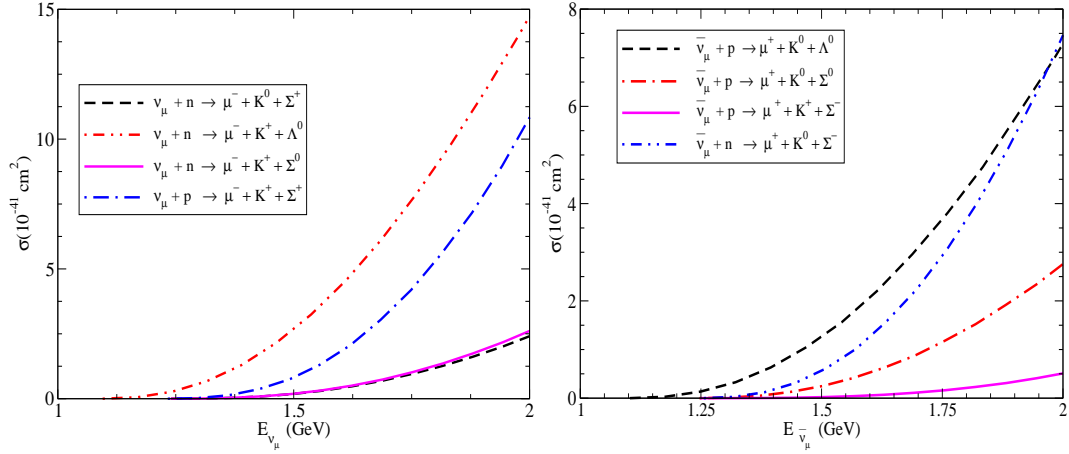
\label{fig:ass_xsec}
 \includegraphics[width=7cm,height=6cm]{associated_nu_v2.eps}
 \includegraphics[width=7cm,height=6cm]{associated_nubar_v2.eps}
 \caption{cross section $|\Delta S|=0$ associated kaon production processes}
\end{figure}

\begin{figure}
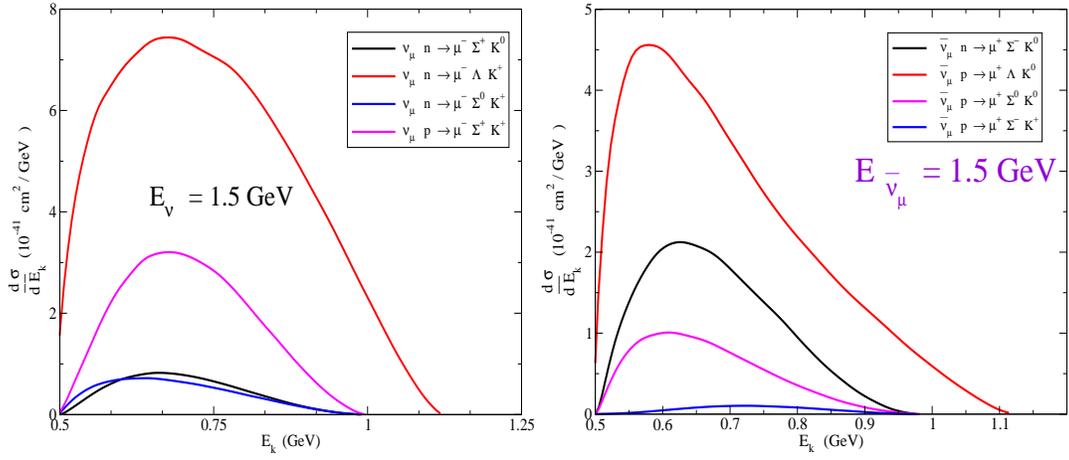
\label{fig:ass_xsec_ek}
 \includegraphics[width=7cm,height=6cm]{nu_Ek_1_5.eps}
 \includegraphics[width=7cm,height=6cm]{nubar_Ek_1_5.eps}
 \caption{Kaon energy distribution for different processes corresponding to neutrino energy of 1.5 GeV.}
\end{figure}

\subsection{Eta production}
It is well known from the studies of real/virtual photon induced $\eta$ production off the nucleon targets that
$\eta$ production is dominated by $S_{11}$(1535) resonance excitation and its subsequent decay in $\eta$. In the case of (anti)neutrino 
induce 1$\eta$ production off a nucleon, the amplitudes corresponding to the nucleon pole and $N^*$
resonances may be written as  

\begin{eqnarray}
J_{N(s)}^\mu &=&  \frac{g V_{ud} }{2\sqrt2} 
\frac{D-3F}{2\sqrt3 f_\pi} \bar u_N (p^\prime) \slashchar{p_\eta} \gamma^5  
\frac{\slashchar{p}+\slashchar{q}+M}{(p+q)^2-M^2} 
{\cal O}^\mu_N u_N (p) \nonumber \\ 
J_{N(u)}^\mu &=&  \frac{g V_{ud} }{2\sqrt2} \frac{D-3F}{2\sqrt3 f_\pi} 
\bar u_N (p^\prime) {\cal O}^\mu_N
  \frac{\slashchar{p}-\slashchar{p_\eta}+M}{(p - p_\eta)^2-M^2} 
\slashchar{p_\eta} \gamma^5 u_N (p), \nonumber \\
J_{R(s)}^\mu & = & \frac{g V_{ud} }{2\sqrt2} i g_{\eta} 
\bar{u}_N (p^\prime)  \slashchar{p_\eta}
 \frac{ \slashchar{p} +\slashchar{q} + M_R }{(p+q)^2 - M_R^2 + i \Gamma_R M_R} 
{\cal O}^\mu_R u_N (p) \nonumber \\
J_{R(u)}^\mu & = &  \frac{g V_{ud} }{2\sqrt2} i g_{\eta} 
\bar{u}_N (p^\prime) {\cal O}^\mu_R
 \frac{ \slashchar{p} -\slashchar{p}_2 + M_R }{(p - p_\eta)^2 - M_R^2 + i \Gamma_R 
M_R} \slashchar{p_\eta} u_N (p), \nonumber \\
\rm{where}&& \nonumber \\ 
{\cal O}^\mu_N &\equiv& f_1^{V}(q^2)\gamma^\mu + f_2^{V}(q^2) i \sigma^{\mu\rho} 
\frac{q_\rho}{2M_N} - f_A (q^2) \gamma^\mu\gamma^5  - f_P (q^2) q^\mu\gamma^5 \nonumber \\
{\cal O}^\mu_R &\equiv&\frac{F_1^{V}(q^2)}{(2 M)^2}(\slashchar{q} q^\mu-q^2\gamma^\mu) \gamma_5
\pm \frac{F_2^{V}(q^2)}{2 M} i \sigma^{\mu\rho} q_\rho \gamma_5   
 - F_A (q^2) \gamma^\mu \mp \frac{F_P (q^2)}{M} q^\mu
\end{eqnarray}
The nucleon form factors $f_{1,2}^V$ are determined in terms of the $f_{1,2}^{p,n}$ 
in the same way as we discussed  in case of $\Delta S=0$ associated strange particle production processes. 
The dipole form is taken for $f_A$ and $f_P$ is related to $f_A$ through PCAC. 
The isovector form factors $F_{1,2}^{V}$ corresponding to the $N^\ast$ resonances,
are given in terms of the electromagnetic transition form factors for  
charged$(F_1^p)$ and neutral$(F_1^n)$ $N^\ast$ resonances:
\begin{equation}
 F_1^V(Q^2) = F_1^ p(Q^2) - F_1^n(Q^2) ;\quad F_2^V(Q^2)= F_2^p(Q^2) - F_2^n(Q^2) .
\end{equation} 
The $F_{1,2}^{p,n}(Q^2)$ are obtained from the helicity amplitudes $A_{\frac{1}{2}}^{p,n}$ and
$S_{\frac{1}{2}}^{p,n}$, given as
% \begin{footnotesize}
\begin{eqnarray}
A_{\frac{1}{2}}^{p,n} &=& \sqrt{\frac{2 \pi \alpha_e }{M}\frac{(M_R+M)^2+Q^2}{M_R^2-M^2}} 
\left( \frac{Q^2}{4M^2} F_1^{p,n}(Q^2) + \frac{M_R-M}{2M} F_2^{p,n}(Q^2) \right) \nonumber \\
S_{\frac{1}{2}}^{p,n} &=& \sqrt{\frac{\pi \alpha_e }{M}\frac{(M_R - M)^2+Q^2}{M_R^2-M^2}} 
\frac{(M_R + M)^2+Q^2}{4 M_R M}
\left(  \frac{M_R-M}{2M} F_1^{p,n}(Q^2) - F_2^{p,n}(Q^2) \right)
\end{eqnarray}
The parameters $A_{\frac{1}{2}}$ and $S_{\frac{1}{2}}$ are generally parameterized as:
\begin{eqnarray*}
 A_{\frac{1}{2}} (Q^2)  &=& A_{\frac{1}{2}} (0) \left(  1 + \alpha \; Q^2 \right) \; e^{ - \beta Q^2} \nonumber \\
 S_{\frac{1}{2}} (Q^2)  &=& S_{\frac{1}{2}} (0) \left(  1 + \alpha \; Q^2 \right) \; e^{ - \beta Q^2} \; ,
\end{eqnarray*}
We fitted the $A_{1/2}^p$ using the data from the MAMI Crystal Ball experiment~\cite{McNicoll:2010qk} and 
for the $A_{1/2}^n$ and $S_{1/2}^{p,n}$ we rely on the latest parameterizations from MAID~\cite{Drechsel:2007if, Tiator:2011pw}.
For resonances $S_{11}(1535)$ and $S_{11}(1650)$ the helicity parameters used for our numerical calculations 
are summarized in Tab.~\ref{tab_eta:hel_amp}. 
\begin{table}[h]
\centering
\caption{Parameters used for the helicity amplitude}
\label{tab_eta:hel_amp}
\begin{tabular}{c c c c c c  c}\hline
Resonance$\rightarrow$&\multicolumn{2}{c}{ S11(1535) } &&S11(1650)&&\\ \hline
Helicity&$A_{\lambda}(0)$&$\alpha$&$\beta$&$A_{\lambda}(0)$ & $\alpha$ & $\beta$\\ 
Amplitude$\downarrow$&$10^{-3}$&&&$10^{-3}$&&\\ \hline
$A_{1/2}^p(Q^2)$&89.38&1.61364&0.75879&53&1.45&0.62 \\ \hline
$S_{1/2}^p(Q^2)$&-16.5&2.8261&0.73735&-3.5&2.88&0.76 \\ \hline
$A_{1/2}^n(Q^2)$&-52.79&2.86297&1.68723&9.3&0.13&1.55 \\ \hline
$S_{1/2}^n(Q^2)$&29.66&0.35874&1.55&10.0&-0.5&1.55\\ \hline
\end{tabular}
\end{table}

\begin{figure}\label{fig:eta_photo_xsec}
   \includegraphics[width=10cm,height=6cm]{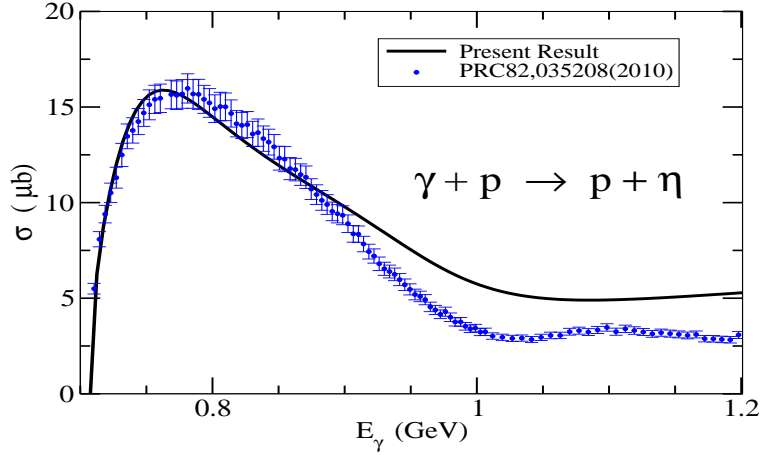}
   \caption{Cross section for photo production of $\eta$-meson. 
   The experimental results are from MAMI Crystal Ball experiment~\cite{McNicoll:2010qk}. }
   \label{photo_eta.eps}
\end{figure}
We derived Goldberger-Treiman relation$(F_A(0) = g_\pi^{N^*})$ and assumed a dipole form for
$ Q^2 $dependence for the axial form factors.
\begin{eqnarray}
F_A ( Q^2 ) &=& F_A (0) \left( 1 + \frac{Q^2}{M_A^2} \right)^{-2}; \\
F_P ( Q^2 ) &=& \frac{( M_R - M ) M}{Q^2 + m_\pi^2} F_A ( Q^2 ).
\end{eqnarray}  
The off-shell effects are taken for the $N^\ast$ resonances by taking 
their total decay width  $\Gamma_R$.
\begin{eqnarray}\label{weight}
 \Gamma_R^{S_{11}(1535)} &=& (.35-.50) \Gamma_{N \pi} + (.42 \pm .10) \Gamma_{N \eta} + (0.01-0.1) \Gamma_{N \pi \pi} \nonumber \\
 \Gamma_R^{S_{11}(1650)} &=& (.5-.90) \Gamma_{N \pi} + (.05-0.15) \Gamma_{N \eta} + (0.03-0.11) \Gamma_{\Lambda K} 
 +  (0.1-0.2) \Gamma_{N \pi \pi},   
\end{eqnarray}
where the partial decay width $\Gamma_{B\phi}$ are calculated using the relation,
\begin{eqnarray}
 \Gamma_{S_{11} \rightarrow B \Phi}={C_\Phi}\left(\frac{g_{\Phi}}{f_\pi}\right)^2\frac{|{\vec p}_{CM}|}{8\pi}
 \frac{(W^2-{M}^2)^2-m_\Phi^2(W^2+{M}^2-2{M}M_R)}{W^2}
\end{eqnarray}
where $C_\Phi=3$ for pion and  $C_\Phi=1$ for eta meson and
\begin{eqnarray}
|{\vec p}_{CM}|=\frac{1}{2W}\sqrt{\left[W^2-(M+m_\Phi)^2\right]~~\left[
W^2-(M-m_\Phi)^2\right]}.
\end{eqnarray} 
$W$ is the energy at resonance rest frame, which for on-mass shell reduces to the 
mass of resonance i.e. $W_{\rm{on-mass}} = M_R$. The quantities in the brackets in Eq.\ref{weight}, are the weightage of partial decay 
widths of the various channel. Comparing the decay width with the available PDG values
enable us to fix the various couplings involved. 

\begin{minipage}{0.45\textwidth}
\begin{eqnarray}
S_{11}^+ (P^\ast) && \nonumber\\
        g_{\pi}^{1650}&=& - 0.105 \nonumber\\
        g_{\eta}^{1650}&=& -0.088 \nonumber\\
        g_{\eta}^{1535}&=&0.284 \nonumber\\
        g_{\pi}^{1535}&=&0.092 \nonumber 
        \end{eqnarray}  
\end{minipage}
\begin{minipage}{0.45\textwidth}
\begin{eqnarray}
 S_{11}^- (N^\ast) && \nonumber\\
        g_{\pi}^{1650}&=& 0.131 \nonumber\\
        g_{\eta}^{1650}&=& 0.0868 \nonumber\\
        g_{\eta}^{1535}&=&0.286 \nonumber\\
        g_{\pi}^{1535}&=&0.106 
\end{eqnarray}  
\end{minipage}

\begin{figure}
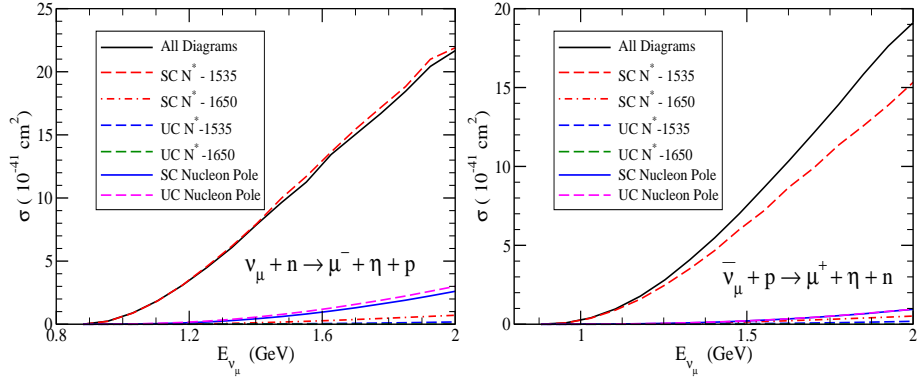
\label{fig:eta_xsec}
 \includegraphics[width=6cm,height=5cm]{Eta_cc_nu_xsec.eps}
 \includegraphics[width=6cm,height=5cm]{Eta_cc_nubar_xsec.eps}
 \caption{Eta production cross section induced by neutrino and antineutrino. 
 Here `SC' and `UC' represents the direct and cross diagrams respectively.}
\end{figure}

\begin{figure}\label{fig:eta_mom}
 \includegraphics[width=6cm,height=5cm]{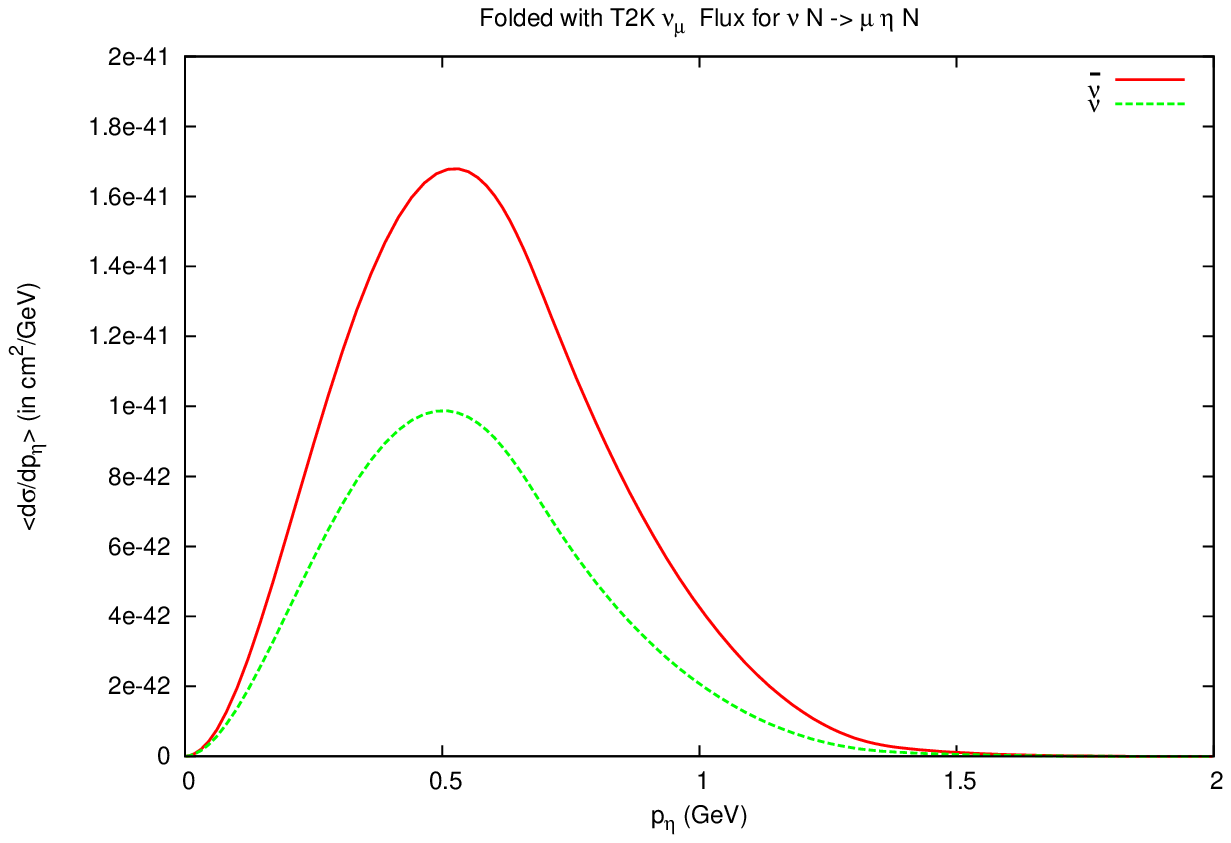}
 \includegraphics[width=6cm,height=5cm]{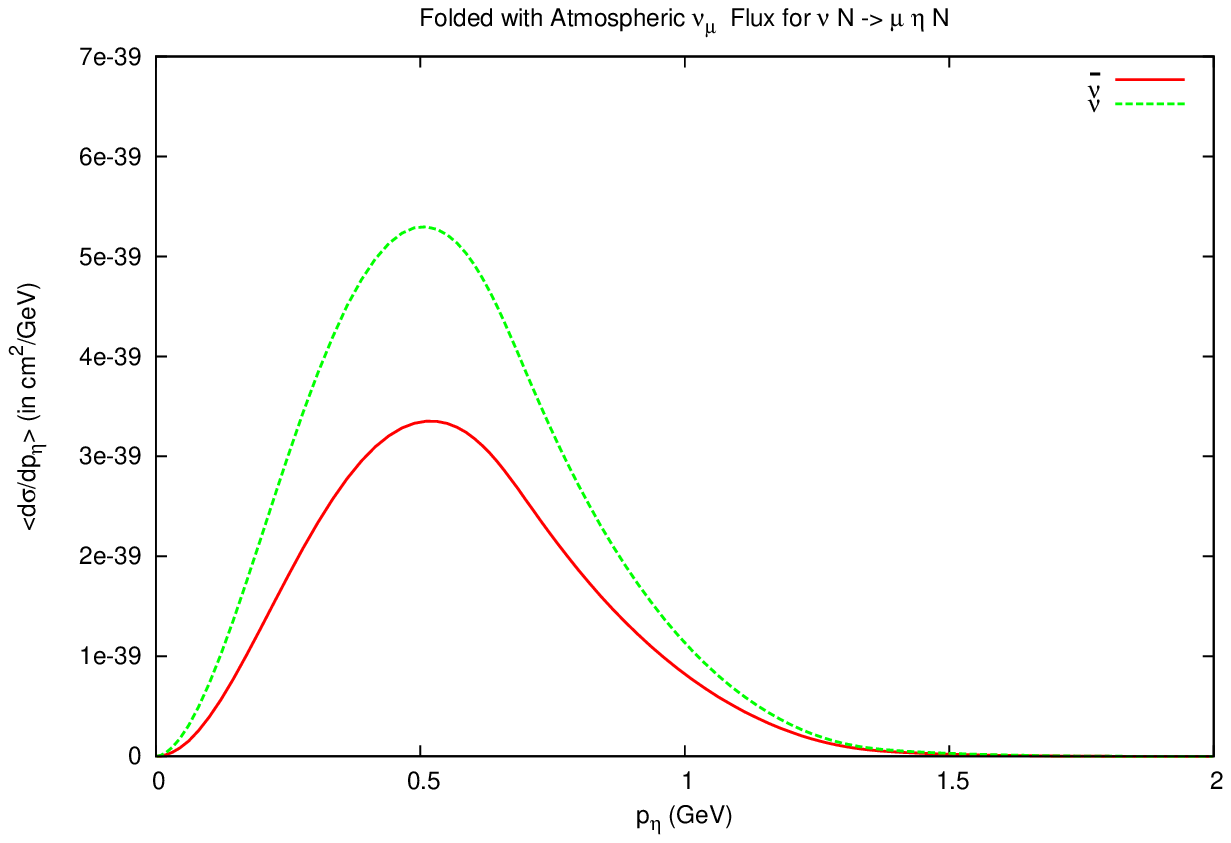}
 \caption{Eta momentum distribution convoluted over T2K $\&$ Atmospheric $\nu$ flux }
\end{figure}

\section{Results and discussions}

The cross sections are obtained after integrating over all the kinematical variables of Eq.~\ref{eq:kinc}.  For $\Delta S$=1 single kaon production processes corresponding to reaction
$\nu_\mu + p \rightarrow  \mu^- + K^+ + p$ and 
$ \bar \nu_\mu + p \rightarrow  \mu^+ + K^- + p$ the results are presented in Fig.~\ref{fig:pp_singleK}.
 It is interesting to note that the cross sections for the neutrino and antineutrino induced reactions are almost the same. 
 In this figure, we have presented the contribution from each term of the Feynman amplitude. We find that the contact term is dominant in both cases.
 In the case of antineutrino induced kaon production the contribution of
$\Sigma^*$ resonance has also been taken into account.
However, the $\Sigma^*$ resonance does not give any significant contribution and 
almost all the contribution comes from the Born terms. We have multiplied the hadronic current by a global dipole form factor 
$f_A(Q^2) = \frac{1}{(1+Q^2/M_d^2)^2}$ with the axial mass $M_d=1$GeV. We have also shown in this figure, the dependence of the cross section on $M_d$.
 The effect of $10\%$ variation on  $M_d$ changes the cross section by 10$\%$ and are 
shown by the shaded area. 

In Fig.~\ref{fig:ass_xsec}, we have presented the results for the total scattering cross section $\sigma$ for the $\Delta S = 0$ 
associated(Y+K) strange particle
production processes for all the reactions shown in Eq.~\ref{Eq1:ass_prod}. 
 In this case also we find(not shown here) that the contribution from the contact term to be dominant one followed by the direct and cross Born term diagrams. 
 Furthermore, we find that the cross sections for the reaction channels with $\Lambda$ in the final state are the largest. 
This can be understood from the relative strength of the 
coupling $g_{NK\Lambda} = \sqrt{3} (D + 3 F)/(6 f_\pi)$ vs $g_{NK\Sigma}=-3 (D-F)/(6 f_\pi)$. 
Apart from the larger coupling, the $\Lambda$ production is favored by 
the available phase space due to its small mass relative to $\Sigma$ baryons. 
For $\nu_\mu n \rightarrow \mu^- \Sigma^+ K^0$ and $\bar \nu_\mu p \rightarrow \mu^+ \Sigma^- K^+ $ 
there is no contribution from the contact term and hence the cross sections are relatively lower.

In Fig.\ref{fig:ass_xsec_ek}, we have presented the results for the kaon energy distribution near threshold at 1.5 GeV for 
(anti)neutrino induced $\Delta S = 0$ processes.
It may be observed that the $\Lambda$ in the final state has the highest peak.
However, the nature of the peaks are little different. For example, in the case of 
neutrino induced process the peak lies around 200MeV of kinetic energy of $\Lambda$ while 
in the case of antineutrino induced process it lies around 50MeV.

 Before doing the numerical calculations for (anti)neutrino induced $\eta$ production off nucleon, first we have obtained the 
 total scattering cross section for photo production of $\eta$-meson and compared them with the experimental results recently obtained 
 from MAMI Crystal Ball experiment~\cite{McNicoll:2010qk}. These results are shown here in Fig.\ref{photo_eta.eps}.
   The vector form factors of the $N-S_{11}$ transition has been obtained using the helicity amplitude
extracted in the analysis of world pion photo- and electro- production data with the unitary isobar model MAID~\cite{Drechsel:2007if, Tiator:2011pw}. 

In Fig.~\ref{fig:eta_xsec}, we have presented the results for neutrino and antineutrino induced $\eta$ production from nucleon
 for the processes given in Eq.~\ref{eta}. Here we have also given the contribution coming from the individual term like s- and u- channel Born diagram 
 and s- and u-channel $S_{11} (1535)$ and $S_{11} (1650)$ resonance contributions. We find that like in the photo- and electro- induced 
 $\eta$ production process, in the case of weak interaction induced process also there is $S_{11}(1535)$ dominance. 
 The  contribution of $S_{11} (1650)$ is negligible.
This can be understood easily as $S_{11}(1535)$ is lighter in mass
and has a relatively larger branching ratio into $\eta N$ than $S_{11}(1650)$. 
The contribution of the non-resonant diagrams in the case of neutrino induced charged current process is 
higher than the corresponding  antineutrino process. We also observe that in the neutrino mode the contribution of 
u-channel diagram is slightly larger than the corresponding s-channel diagram.
Furthermore, we have studied flux integrated $p_\eta$ dependence corresponding to 
(anti)neutrino flux at T2K and atmospheric neutrinos. The results are presented in Fig.~\ref{fig:eta_mom}.

To conclude, in this work we have studied single kaon/antikaon, eta and associated particle production processes in the (anti)neutrino
 induced reactions from nucleon targets. These results may be helpful in the analysis of MINER$\nu$A experiment as well as the experiments looking for proton decay.
\begin{theacknowledgments}
One of the authors(M Rafi Alam) is thankful to the organizers of the CETUP workshop for the warm hospitality.
\end{theacknowledgments}

\bibliographystyle{aipproc}   % if natbib is available 

\end{document}